\documentclass[twocolumn, amssymb, nobibnotes, aps, prl, superscriptaddress, 10pt]{revtex4-2}
\usepackage{bbm}
\pdfoutput=1
\usepackage[utf8]{inputenc}
\usepackage[english]{babel}
\usepackage[T1]{fontenc}
\usepackage{bm}
\usepackage{mathtools}
\usepackage{amssymb}
\usepackage{amsthm}
\usepackage{physics}
\usepackage{braket}
\usepackage[colorlinks=True, linkcolor= red, citecolor= blue]{hyperref}
\usepackage{nameref}
\usepackage{enumitem}
\usepackage{xcolor}

\def\Ex#1{\left\langle {#1} \right\rangle}

\newcommand{\sig}{b}
\newcommand{\est}{\sig_{\text{est}}}
\newcommand{\bw}{\Omega}

\begin{document}


\title{Entanglement-enhanced Ramsey spectroscopy of broadband signals}

\author{Noah Kaufmann}
\affiliation{Center for Hybrid Quantum Networks (Hy-Q), The Niels Bohr Institute, University of Copenhagen,  Jagtvej 155A, Copenhagen DK-2200, Denmark}

\author{Anders Søndberg Sørensen}
\affiliation{Center for Hybrid Quantum Networks (Hy-Q), The Niels Bohr Institute, University of Copenhagen,  Jagtvej 155A, Copenhagen DK-2200, Denmark}

\begin{abstract}
    The use of entanglement in sensing is often motivated by improved sensitivity arising from non-classical correlations among probes. For static signals, however, this quantum advantage in Ramsey spectroscopy with $N$ probes is known to be highly susceptible to dephasing, which severely limits the possible gain from entanglement. In contrast,  we consider sensing of signals that vary in time. Considering the simplest example of dephasing by classical white noise, we show that entanglement allows a large increase in bandwidth when estimating the frequency components of a time-dependent signal. Without sacrificing any sensitivity, entanglement allows for an increase in bandwidth by a factor of $N$. Alternatively, when interrogation times are limited by the signal bandwidth rather than by dephasing, the average estimation uncertainty over a fixed bandwidth is reduced by a factor of $\sqrt{N}$ compared with separable states.
\end{abstract}

\maketitle

\textit{Introduction.--} The fundamental task of sensing is to obtain information about a signal through its interaction with probes whose preparation and measurement can be controlled~\cite{Giovannetti2006}. One of the main goals of quantum technologies is to exploit quantum correlations to improve the sensitivity in such setups. A true quantum advantage arises when, for a fixed number of probes $N$ and total interaction time $T$, a sensing protocol extracts more information about the signal $\sig(t)$ than any protocol restricted to separable control of the probes~\cite{Degen2017}. Prominently studied is the uncertainty scaling in estimating a static signal with entangled atomic ensembles~\cite{Kitagawa1993, Wineland1994, Bollinger1996, Huelga1997, Matsuzaki2011, Chin2012}, where, in principle, entanglement enables the sensitivity of $N$ probes to scale linearly with $N$, in contrast to the $\sqrt{N}$ scaling achievable with separable states~\cite{Bollinger1996, Giovannetti2006}. However, this quantum advantage is highly susceptible to noise, and the enhanced scaling of the sensitivity almost vanishes if the probes are affected by uncorrelated dephasing~\cite{Huelga1997, Escher2011, Demkowicz2012}. 

So far, most studies of quantum sensing have aimed at estimating static parameters. In this Letter, we show that a more noise-resilient entanglement enhancement arises when trying to sense a time-varying signal. This is a practically relevant scenario if the aim is to resolve the frequency components $\sig(\omega)$ of a time-dependent signal with bandwidth $\bw$. According to the Nyquist-Shannon sampling theorem, the signal must be sampled at a frequency greater than twice the highest signal frequency~\cite{Oppenheim1997}. This limits the interrogation time of the probes, and therefore the effect of decoherence, allowing a significant enhancement from entanglement in the average sensitivity across the bandwidth. 

\begin{figure*}[ht!]
  \centering
    \includegraphics[width=\textwidth]{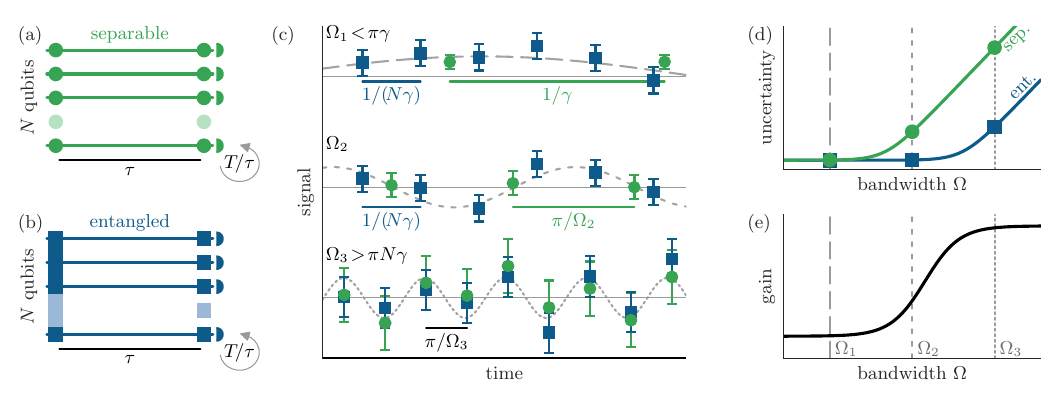}
  \caption{Ramsey spectroscopy of time-varying signals in the frequency range $[0, \Omega]$ under Markovian dephasing of strength $\gamma$, using either $N$ separable (a) or entangled (b) probes. The interrogation time $\tau$ is the free-evolution time during which the signal is encoded in the probes, and the protocol is repeated $T/\tau$ times. For the entangled probes, the preparation involves interaction among the probes. The probes do not interact during evolution and probing. (c) Qualitative optimal interrogation times and corresponding uncertainties for signals at the maximum frequency $\Omega_i$ of the band $[0, \Omega_i]$. For low bandwidth $\Omega_1 \!<\! \pi \gamma$, $\tau$ is limited by dephasing, leading to shorter interrogation times for entangled probes. At high bandwidth $\Omega_3 \!>\! \pi N \gamma$, where the Nyquist frequency fixes $\tau$ for both protocols, entangled probes have higher sensitivity due to the enhanced signal. In the intermediate regime $ \pi \gamma \!<\! \Omega_2 \!<\! \pi N \gamma$, separable probes are Nyquist-limited, whereas entangled probes remain dephasing-limited. (d) Resulting estimation uncertainty of the highest-frequency component as a function of bandwidth. (e) Metrological gain, defined as the uncertainty ratio between the separable and the entangled protocol. Panels (d,e) are shown on log-log scales.}
  \label{fig:fig1}
\end{figure*}

To illustrate this, we consider $N$ two-level systems with a signal-dependent energy splitting~\cite{Bollinger1996}. In a Markovian noise environment, preparing the $N$ probes in a maximally entangled GHZ state is equivalent to creating a single probe with an $N$-fold enhanced signal-sensitivity, but an $N$-fold reduced dephasing time. For static-signal estimation, these effects exactly compensate: the enhanced signal accumulation is canceled by the equally enhanced coupling to dephasing noise, yielding the same estimation uncertainty as for separable probes~\cite{Huelga1997}. Importantly, however, the increased response to both signal and noise leads to the sampling rate of GHZ probes being $N$ times higher than for separable probes~\cite {Huelga1997}. Consequently, GHZ probes naturally provide a broader bandwidth than separable probes, as illustrated schematically in Fig.~\ref{fig:fig1}. 

This increase in bandwidth has been identified as a benefit of entanglement-enhanced sensing~\cite{Huelga1997, Xia2023, Ye2024, Boyers2025}, including in AC magnetometry under Markovian noise~\cite{Sichanugrist2025}, while Fisher-information-based bounds have been derived for sensitivity--bandwidth tradeoffs and broadband detection~\cite{Wan2022, Polloreno2023, polloreno2026}. More broadly, increased bandwidth can improve atomic-clock stability~\cite{Andre2004, Kessler2014, Colombo2022} and is relevant for many fundamental physics experiments~\cite{Pospelov2013, Smorra2019, Sedda2020, Stadnik2020}. So far, however, a detailed analysis of bandwidth enhancement using entangled probes in Ramsey spectroscopy under realistic dephasing is lacking. In this Letter, we provide such an analysis for an ensemble of two-level systems, focusing on GHZ and spin-squeezed states and treating both Markovian and non-Markovian dephasing. We show that, in contrast to the sensitivity enhancement for static signals, the entanglement-induced increase in bandwidth is substantially more resilient to dephasing, making it a practically viable quantum advantage. Moreover, by formulating a protocol-independent signal-to-noise type argument, we derive a bound on the average estimation uncertainty over a fixed bandwidth that depends solely on the noise power spectral density.

\textit{Setting.--} We consider a real-valued, time-dependent external field $\sig(t)$ of bandwidth $\bw$. We discuss the estimation of its frequency components $\sig(\omega)$ using $N$ identical qubits whose energy splittings are coupled to the signal. The dynamics in the presence of the external field and a classical stationary pure-dephasing environment are described by the effective Hamiltonian ($\hbar = 1$)~\cite{Bing2019}\vspace{-0.3cm}
\begin{equation}
\begin{aligned}
    \label{eq:system_hamiltonian}
    H(t) =& \frac{1}{2} \sum_{n=1}^N \left[\sig(t) + F_n(t) \right]\sigma_z^{(n)}, \\
    =& \frac{1}{2} \sum_{n=1}^N \int_{-\infty}^\infty e^{i \omega t} \left[\Tilde{\sig}(\omega) + \Tilde{F}_n(\omega) \right] d\omega \,\, \sigma_z^{(n)},
\end{aligned}
\end{equation}
where $\sigma_z^{(n)}$ is the $z$-Pauli operator of qubit $n$ with internal states $\ket{0}_n$ and $\ket{1}_n$, and the noise is modeled by real stochastic processes $F_n(t)$ of mean zero. We assume that the noise of each particle arises from an independent, identically distributed Gaussian process, i.e., the noise is uncorrelated among particles.

In the second line of Eq.~\eqref{eq:system_hamiltonian} we express everything in terms of Fourier components at frequency $\omega$, which is the natural representation for stationary processes. We note here that we consider a scenario where we allow temporal correlations between different shots of the experiment~\cite{Kaufmann2026, Gosh2025} such that the noise can be expressed through its power spectral density $2 \pi \Ex{\Tilde{F}_n(\omega) \Tilde{F}_{n'}^*(\omega')} = \delta_{n n'} \delta(\omega - \omega')  S_F(\omega)$, where we have further assumed that the probes interact with independent baths.

For a finite measurement time $T$, the frequency resolution is \(2\pi/T\), and we define $\sig(\omega_k) = 1/T \int_0^T \sig(t) e^{-i\omega_kt} dt$ with $\omega_k = 2\pi k / T$ and analogously for $F_n(\omega_k)$. Each estimate $\est(\omega_k)$ obtained by the sensing protocol is connected to a frequency-dependent uncertainty $\Delta \est (\omega_k)$. We consider unbiased estimators $\Ex{\est(\omega_k)} = \sig(\omega_k)$, such that $[\Delta \est(\omega_k)]^2 = \Ex{\left| \sig(\omega_k) - \est(\omega_k) \right|^2}$ is the estimator's mean square error. For a given signal bandwidth $\bw$, we compare different sensing protocols according to their average uncertainty, \vspace{-0.1cm}
\begin{equation}
\label{eq:figure_of_merit}
    \Delta B(\bw) =\int_{0}^{\bw} \Delta \sig_\text{est}(\omega) \, \, \frac{d\omega}{\bw},
\end{equation}
where in the limit of $T\rightarrow \infty$ we replaced the sum over $\omega_k$ by an integral. In particular, we are interested in the metrological gain $r$ achieved using entanglement compared to separable strategies $r = \Delta B^\text{sep}(\bw) / \Delta B^\text{ent}(\bw)$.


\textit{Fundamental limit.--} Following the approach of Ref.~\cite{Kaufmann2026}, we derive a protocol-independent fundamental limit on the average uncertainty~\eqref{eq:figure_of_merit} based on the noise power spectral density in the asymptotic limit $T \rightarrow \infty$. Because both the signal $\sig(t)$ and the noise $F_n(t)$ couple to the dynamics through the same operator $\sigma_z$, the probe cannot distinguish a signal component $\sig(\omega)$ from the collective noise at that frequency. Consequently, the variance of the average noise at some frequency $\omega$ provides a lower bound for the estimation uncertainty $\Delta \est(\omega)$. In short, the sensitivity at a given frequency is limited by the probe’s signal-to-noise ratio at that frequency.

This intuition can be formalized by writing $H(\omega) = \frac{1}{2}\left[\sig(\omega) + \Bar{F}(\omega)\right] \sum_n \sigma_z^{(n)} + \frac{1}{2} \sum_n \delta F_n (\omega) \sigma_z^{(n)}$, where $\Bar{F}(\omega) = N^{-1} \sum_n F_n(\omega)$ is the average noise and $\delta F_n(\omega) = F_n(\omega) - \Bar{F}(\omega)$ its deviation for the $n$th probe. $\Bar{F}(\omega)$ and $\sig(\omega)$ enter the dynamics equivalently, implying for independent and identically distributed noise across probes the bound $\Delta \est(\omega) \geq \Delta \Bar{F}(\omega) = \left[\Delta F_1(\omega)\right] / \sqrt{N}$.

For $T$ much longer than the noise correlation time, $[\Delta F_1(\omega)]^2=S_F(\omega)/T$, and replacing the frequency sum by an integral gives~\cite{Ziemer2014}\vspace{-0.1cm}
\begin{equation}
\label{eq:fundamental-bound}
    \Delta B (\bw) \geq \frac{1}{\sqrt{NT}} \int_0^{\bw} \sqrt{S_F(\nu)} \, \, \frac{d\nu}{\bw}.
\end{equation}
The derivation includes only dephasing noise, neglecting shot-noise contributions, hence the bound becomes trivial and does not recover the Heisenberg scaling for $S_F(\nu) = 0$. Nevertheless, it sets a lower bound on the achievable  sensitivity in the presence of dephasing.


\textit{Ramsey spectroscopy.--} The canonical approach to sensing with two-level systems is Ramsey spectroscopy~\cite{Degen2017}. In each Ramsey shot, illustrated in Fig.~\hyperref[fig:fig1]{\ref{fig:fig1}a}, an initial state is prepared, evolves freely according to $H(t)$ for time $\tau$, and is measured with observable $M_l$. Repeating the protocol $L=T/\tau$ times yields outcomes $\{m_l\}_{l=0}^{L-1}$, from which $\sig(\omega)$ is estimated. We assume that the time needed for state preparation and measurements is much shorter than the interrogation time. The interrogation time is fixed over all repetitions, and the total experimental time $T$ is much longer than both the noise correlation time and the interrogation time. 

The analysis is restricted to weak signals $\sig(t)$, where the accumulated phase $\varphi_l$ during the free evolution is small $|\varphi_l| = |\int_{l\tau}^{l\tau + \tau} \sig(t) \, dt | \ll 1 / N$. In this regime and with measurements $M_l$ satisfying $\Ex{M_l} = 0$ for $\varphi_l = 0$, the phase $\varphi_l$ can be estimated from the measurement outcome $m_l$ with the linear estimator $(\varphi_l)_{\text{est}} = m_l / \left(\partial \Ex{M_l} / \partial \varphi_l\vert_{\varphi_l=0}\right)$~\cite{Pezze2018}. The frequency components of the signal are then estimated by Fourier transforming $\{(\varphi_l)_\text{est}\}_{l=0}^{L-1}$:
\begin{equation}
    \est(\omega_k) = \frac{1}{T \, \text{sinc}\left(\frac{\omega_k \tau}{2}\right)}  \sum_{l = 0}^{L-1} \! \frac{m_l}{\frac{\partial \Ex{M_l}}{\partial \varphi_l}\!\bigr|_{\varphi_l=0}} e^{-i \omega_k \tau (l +\frac{1}{2})}.
\end{equation}
The prefactor $\text{sinc}(\omega_k \tau/2)^{-1} \!\geq\! 1$ originates from the fact that the experiment does not probe the signal but the integrated signal, which weakens the signal from higher frequency components compared to lower frequencies. The response of the measurement to the signal, $\partial \Ex{M_l} / \partial \varphi_l$, is independent of $l$ as we consider stationary noise. However, it depends on the noise, whose statistics are assumed to be determined during sensor calibration. The uncertainty $\Delta \est(\omega_k)$ is obtained from the correlations $\Ex{M_l M_{l'}}$, calculated by evolving the measurement operators in the Heisenberg picture, evaluating them in the prepared state $\ket{\psi_0}$, and averaging over noise realizations.

\textit{Separable probe states.--} We benchmark the entangled protocols against standard Ramsey spectroscopy~\cite{Degen2017}. Here, each qubit is first rotated from the ground state to a superposition state by a $\pi/2$ pulse around the spin $y$-axis, $\ket{\psi_0^{\text{sep}}} = \ket{+}^{\otimes N}$, where $\ket{+} = (\ket{0} + \ket{1})/\sqrt{2}$. After the free evolution, the measurement consists of a $\pi/2$ pulse around the spin $x$-axis and a measurement in the $z$-basis, hence $M^{\text{sep}} = J_y$.

\textit{Spin-squeezed probe states.--} Spin-squeezed states can improve the sensitivity by reducing the shot noise~\cite{Kitagawa1993, Wineland1994, Boyers2025}. The protocol is the same as with separable input states, hence $M^{\text{sss}} \!=\! J_y$, but the initial state is squeezed to reduce $\Delta J_y$. We approximate the optimal spin-squeezed states~\cite{Sorensen2001} with the Gaussian states of Ref.~\cite{Andre2004, Borregard2013}. 

\textit{GHZ probe states.--} GHZ states are ideal for sensing in noiseless environments~\cite{Bollinger1996}. For estimating static signals in the presence of dephasing, their optimal interrogation time scales with $1/N$~\cite{Huelga1997}, which makes them interesting for sensing across a broad bandwidth. The initial state is $\ket{\psi_0^{\text{GHZ}}} = 1/\sqrt{2}(\ket{0}^{\otimes N} + \ket{1}^{\otimes N})$ and the evolved state is measured by projection onto $1/\sqrt{2}(\ket{0}^{\otimes N} - i \ket{1}^{\otimes N})$, with $1/2$ being subtracted such that $\Ex{M_l} = 0$ for $\varphi_l = 0$ is fulfilled. The corresponding measurement can also be implemented by single-qubit operations and a separable measurement~\cite{Bollinger1996}.

\textit{No noise.--} In the absence of noise, $F_n(t)=0$, it is optimal to maximize the interrogation time to minimize shot noise.  Hence, the optimal sampling rate is given by the Nyquist frequency, $\tau_{\text{opt}} = \pi/\bw$, leading to
\begin{equation}
\label{eq:NoNoise}
    \Delta B^{\text{sss}} \! = \! \sqrt{\!\frac{\Ex{J_y^2} \bw}{\Ex{J_x}^2 T \pi}} \frac{4 C}{\pi}, \quad \Delta B^{\text{GHZ}} \! = \! \sqrt{\!\frac{\bw}{N^2 T \pi}} \frac{4 C}{\pi},
\end{equation}
for spin-squeezed and GHZ states, respectively. Here $C$ is the Catalan's constant, and the factor $4 C / \pi \approx 1.16$ comes from integrating $\text{sinc}(\omega \tau / 2)^{-1} d\omega$ from 0 to $\pi / \tau$. For the considered separable protocol we have $\Ex{J_y^2} / \Ex{J_x}^2 = 1 / N$ in $\Delta B^{\text{sss}}$, and hence $\Delta B^{\text{sep}} \propto \sqrt{\bw / N}$. GHZ states lead to more precise estimates scaling as $\Delta B \propto \sqrt{\bw} /N$. Optimally spin-squeezed states satisfy $\Ex{J_y^2} / \Ex{J_x}^2 \propto 1 / N^2$, thus reaching the same scaling as GHZ states. The metrological gain $r \propto \sqrt{N}$ is independent of the bandwidth, similar to the discussion of sensing of static signals~\cite{Wineland1994, Bollinger1996}.

\begin{figure*}[t]
\centering
\includegraphics[width=\textwidth]{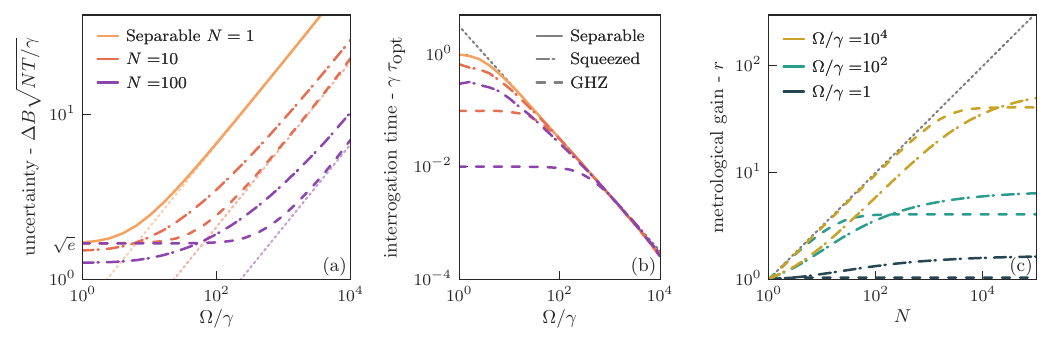}
  \caption{Bandwidth-dependent sensing uncertainty with separable, squeezed and GHZ probe states in the presence of Markovian dephasing of strength $\gamma$. (a) The average estimation uncertainty as a function of the signal bandwidth for different numbers of probes. (b) The interrogation time that minimizes the average estimation uncertainty at a given signal bandwidth is plotted against the bandwidth. (c) The metrological gain $r$ versus the number of probes for different bandwidths. Solid lines are for separable states, while dashed (dashed-dotted) lines are for GHZ (squeezed states). Colors in (a) and (b) denote the number of probes (N), whereas colors in (c) describe the ratio of bandwidth to dephasing  $\Omega/\gamma$. The dotted lines correspond to GHZ probes and no noise. }
  \label{fig:fig2}
\end{figure*}

\textit{Markovian dephasing.--} The discussion gets more interesting for Markovian dephasing characterized by a strength $\gamma$, defined by $\Ex{F_n(t) F_{n'}(t')} = \gamma \delta(t - t') \delta_{n,n'}$. In this case, the stochastic Hamiltonian formalism of Eq.~\eqref{eq:system_hamiltonian} corresponds to a master equation with jump operators $L_n =\sqrt{\gamma} \sigma_z^{(n)}/2$ and system Hamiltonian $H = \sig(t) \, J_z$. We obtain the following average uncertainty,
\begin{equation}
\label{eq:Markovian}
\begin{aligned}
    &\Delta B_\tau^{\text{sep}} = \sqrt{\frac{e^{\gamma \tau}}{N T \tau}}  \!\int_0^{\bw} \!\! \text{sinc}\left(\!\frac{\omega \tau}{2}\!\right)^{\!-1} \frac{d \omega}{\bw},\\
    &\Delta B_\tau^{\text{sss}} = \sqrt{\frac{\frac{N}{4}\left( e^{\gamma \tau} \!-\!1\right)\!+\!\Ex{J_y^2}}{\Ex{J_x}^2 T \tau}} \!\int_0^{\bw} \!\! \text{sinc}\left(\!\frac{\omega \tau}{2}\!\right)^{\!-1} \!\!\frac{d \omega}{\bw},\\
    &\Delta B_\tau^{\text{GHZ}} = \sqrt{\frac{e^{N \gamma \tau}}{N^2 T \tau}}  \!\int_0^{\bw} \!\! \text{sinc}\left(\!\frac{\omega \tau}{2}\!\right)^{\!-1} \frac{d \omega}{\bw}.
    \end{aligned}
\end{equation} 

Fig.~\ref{fig:fig2} shows the results obtained by optimizing the interrogation time and squeezing to minimize the average uncertainty in Eq.~\eqref{eq:Markovian}, for different probe numbers $N$ and signal bandwidths $\bw$. Three regimes emerge, corresponding to whether the separable and entangled protocols are dephasing- or Nyquist-limited, as sketched in Fig.~\hyperref[fig:fig1]{\ref{fig:fig1}c}. In the dephasing-limited regime, the optimal interrogation time is fixed by the noise and therefore independent of bandwidth, while in the Nyquist-limited regime it scales inversely with bandwidth, as shown in Fig.~\hyperref[fig:fig2]{\ref{fig:fig2}b}. For separable probes, the corresponding crossover bandwidth is independent of $N$, $\Tilde{\bw}_{\text{sep}} \approx \pi\gamma$. For spin-squeezed and GHZ states, the crossover is shifted to larger bandwidths, scaling as $\Tilde{\bw}_{\text{sss}} \approx \pi N^{1/3}\gamma$ and $\Tilde{\bw}_{\text{GHZ}} \approx \pi N\gamma$, respectively. The crossovers are smooth, as seen in Fig.~\hyperref[fig:fig2]{\ref{fig:fig2}b}, because the $\mathrm{sinc}(\omega\tau/2)^{-1}$ factor already favors shorter interrogation times before the Nyquist limit is reached.

In the dephasing-limited regime, where $\gamma > \bw/\pi$ (see $\Omega_1$ in Fig.~\hyperref[fig:fig1]{\ref{fig:fig1}c}), the behavior reduces to that of sensing a static signal, and $\Delta B(\bw)$ becomes bandwidth independent, as seen in Fig.~\hyperref[fig:fig2]{\ref{fig:fig2}a} for $\bw < \Tilde{\bw}_{\text{sep}}$. For $N\gg1$,  spin-squeezed probes outperform GHZ and separable probes by a factor of $\sqrt{e}$ (see $\bw / \gamma = 1$ in Fig.~\hyperref[fig:fig2]{\ref{fig:fig2}c})~\cite{Huelga1997}.

In the regime $\bw > \Tilde{\bw}_{\text{GHZ}}$, where the interrogation times of both entangled and separable probes are Nyquist-limited (see $\Omega_3$ in Fig.~\hyperref[fig:fig1]{\ref{fig:fig1}c}), the shot-noise contribution to the estimation uncertainty dominates over dephasing. As a result, the no-noise behavior of Eq.~\eqref{eq:NoNoise} is recovered, with $\Delta B (\bw) \propto \sqrt{\bw}$, as shown in Fig.~\hyperref[fig:fig2]{\ref{fig:fig2}a} for $\bw > \Tilde{\bw}_{\text{GHZ}}$. This leads to $r \propto \sqrt{N}$~\cite{Bollinger1996, Pezze2018, Colombo2022}, as visible in Fig.~\hyperref[fig:fig2]{\ref{fig:fig2}c} at small probe numbers for $\bw/\gamma = 10^2$ and $10^4$.

Finally, for $\pi \gamma \!\approx\! \Tilde{\bw}_{\text{sep}} \!<\! \bw \!<\! \Tilde{\bw}_{\text{GHZ}} \!\approx\! \pi N \gamma$, respectively $\Tilde{\bw}_{\text{sep}} \!<\! \bw \!<\! \Tilde{\bw}_{\text{sss}} \!\approx\! \pi N^{1/3} \gamma$, the interrogation time of the separable probes is Nyquist-limited, while it is dephasing-limited for entangled probes (see $\Omega_2$ in Fig.~\hyperref[fig:fig1]{\ref{fig:fig1}c}). In this intermediate regime, the metrological gain is independent of $N$ and scales as $\sqrt{\bw}$. As the entangled probes are dephasing-limited, the spin-squeezed states achieve a gain greater by a factor of $\sqrt{e}$ compared to the GHZ probes for large $N$. This is visible in Fig.~\hyperref[fig:fig2]{\ref{fig:fig2}c} for large probe numbers for $\bw/\gamma \!=\! 10^2$ and $10^4$, where $r \propto \sqrt{\Omega/\gamma}$.

\begin{figure}[]
\centering
\includegraphics[width=\columnwidth]{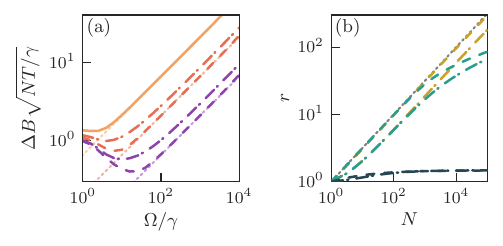}
  \caption{The average uncertainty and metrological gain 
  of spin-squeezed and GHZ probes are plotted in the presence of noise with spectral density $S_F(\nu) = \gamma e^{-\nu^2 / (2 \sigma^2)}$ for $\gamma = \sigma$. The same parameters, colors, and line styles as in Fig.~\ref{fig:fig2} are used. (a) The normalized average uncertainty versus the signal bandwidth $\bw$ for different numbers of probes $N$. (b) The corresponding metrological gain as a function of $N$ for different $\bw$.}
  \label{fig:fig3}
\end{figure}

\vspace{0.15cm}\textit{Non-Markovian dephasing.--} For temporally correlated classical dephasing noise, the accumulated dephasing does not build up linearly in time.  Here, we focus on correlated noise with a Gaussian spectrum $S_F(\nu) = \gamma e^{- \nu^2 / (2 \sigma^2)}$, which highlights the qualitative differences from the Markovian case.

Fig.~\ref{fig:fig3} shows the analog of Figs.~\hyperref[fig:fig2]{\ref{fig:fig2}a} and~\hyperref[fig:fig2]{\ref{fig:fig2}c} for correlated noise. Since the considered noise model peaks at low frequencies, the estimation noise drops as we increase the bandwidth because we probe frequency components with lower noise. In the Nyquist-limited regime, visible at large bandwidths in Fig.~\hyperref[fig:fig3]{\ref{fig:fig3}a}, the average estimation uncertainty is shot-noise dominated and follows the Markovian and noiseless behavior. In contrast to Markovian noise, GHZ probes outperform separable probes in the dephasing-limited regime, because the accumulated phase noise grows quadratically rather than linearly in time for short $\tau$~\cite{Matsuzaki2011, Chin2012, Kaufmann2026}.

For a fixed signal bandwidth, the metrological gain is maximal in the mixed regime, which is reached for high N in Fig.~\hyperref[fig:fig3]{\ref{fig:fig3}b}. In this regime, the Nyquist-limited separable protocol remains shot-noise dominated and can therefore be approximated by the noiseless expression of Eq.~\eqref{eq:NoNoise}. Meanwhile, the entangled protocol is dephasing-limited. Assuming that, for increasing $N$, its uncertainty approaches the fundamental bound in Eq.~\eqref{eq:fundamental-bound}, we obtain the following approximate expression for the maximal metrological gain $r_{\text{max}}$,
\begin{equation}
    r_{\text{max}} (\bw) \propto \frac{\bw^{3/2}}{\int_0^{\bw} \sqrt{ S_F(\nu)} d\nu}.
\end{equation}
In contrast to the Markovian case, where $r_\text{max} \propto \bw^{1/2}$, finite-bandwidth noise yields $r_\text{max} \propto \bw^{3/2}$ once $\bw$ exceeds the noise bandwidth (compare Fig.~\hyperref[fig:fig2]{\ref{fig:fig2}c} to  Fig.~\hyperref[fig:fig3]{\ref{fig:fig3}b}).


\textit{Discussion.--} We have shown that entanglement enhances the sensing of signals with a broad bandwidth. In particular, there is a $\sqrt{N}$ improvement in the sensitivity even in the presence of Markovian dephasing if the required bandwidth exceeds approximately $N$ times the dephasing rate. For a given signal bandwidth, the maximal metrological gain scales with $\sqrt{\bw}$ for Markovian and $\bw^{3/2}$ for the considered non-Markovian noise. These results highlight that while the  gain in sensitivity may not be so prominent when trying to sense static signals, there is a clear advantage from entanglement  when considering time-varying signals of high bandwidth. An interesting extension of this work would be to relax the condition of a fixed interrogation time, to consider adaptive, nondestructive, weak or entangled measurements~\cite{Shah2010, Gosh2025}, and to investigate different probe state preparations. From the  crossing    of the sensitivity of the spin-squeezed and maximally entangled state at a fixed bandwidth in Fig.~\hyperref[fig:fig2]{\ref{fig:fig2}a}, it seems plausible that a more general class of states could interpolate between the two regimes and have a better performance in the intermediate regime.

\textit{Acknowledgments.--} We thank  Kasper H. Nielsen and Eva M. González-Ruiz for their contributions to this project. This work was supported by Danmarks Grundforskningsfond (Grant No. DNRF139, Hy-Q Center for Hybrid Quantum Networks).

\bibliography{References}

@article{Kitagawa1993,
  title = {Squeezed spin states},
  author = {Kitagawa, Masahiro and Ueda, Masahito},
  journal = {Phys. Rev. A},
  volume = {47},
  issue = {6},
  pages = {5138--5143},
  numpages = {0},
  year = {1993},
  month = {Jun},
  publisher = {American Physical Society},
  doi = {10.1103/PhysRevA.47.5138},
  url = {https://link.aps.org/doi/10.1103/PhysRevA.47.5138}
}

@article{Wineland1994,
  title = {Squeezed atomic states and projection noise in spectroscopy},
  author = {Wineland, D. J. and Bollinger, J. J. and Itano, W. M. and Heinzen, D. J.},
  journal = {Phys. Rev. A},
  volume = {50},
  issue = {1},
  pages = {67--88},
  numpages = {0},
  year = {1994},
  publisher = {American Physical Society},
  doi = {10.1103/PhysRevA.50.67},
  url = {https://link.aps.org/doi/10.1103/PhysRevA.50.67}
}

@article{Bollinger1996,
  title = {Optimal frequency measurements with maximally correlated states},
  author = {Bollinger, J. J. and Itano, Wayne M. and Wineland, D. J. and Heinzen, D. J.},
  journal = {Phys. Rev. A},
  volume = {54},
  issue = {6},
  pages = {R4649--R4652},
  numpages = {0},
  year = {1996},
  month = {Dec},
  publisher = {American Physical Society},
  doi = {10.1103/PhysRevA.54.R4649},
  url = {https://link.aps.org/doi/10.1103/PhysRevA.54.R4649}
}

@book{Oppenheim1997,
  title={Signals \& systems},
  author={Oppenheim, Alan V and Willsky, Alan S and Nawab, Syed Hamid},
  year={1997},
  publisher={Pearson Educaci{\'o}n}
}

@article{Huelga1997,
  title = {Improvement of Frequency Standards with Quantum Entanglement},
  author = {Huelga, S. F. and Macchiavello, C. and Pellizzari, T. and Ekert, A. K. and Plenio, M. B. and Cirac, J. I.},
  journal = {Phys. Rev. Lett.},
  volume = {79},
  issue = {20},
  pages = {3865--3868},
  numpages = {0},
  year = {1997},
  month = {Nov},
  publisher = {American Physical Society},
  doi = {10.1103/PhysRevLett.79.3865},
  url = {https://link.aps.org/doi/10.1103/PhysRevLett.79.3865}
}

@article{Sorensen2001,
  title = {Entanglement and Extreme Spin Squeezing},
  author = {S\o{}rensen, Anders S. and M\o{}lmer, Klaus},
  journal = {Phys. Rev. Lett.},
  volume = {86},
  issue = {20},
  pages = {4431--4434},
  numpages = {0},
  year = {2001},
  month = {May},
  publisher = {American Physical Society},
  doi = {10.1103/PhysRevLett.86.4431},
  url = {https://link.aps.org/doi/10.1103/PhysRevLett.86.4431}
}

@article{Andre2004,
  title = {Stability of Atomic Clocks Based on Entangled Atoms},
  author = {Andr\'e, A. and S\o{}rensen, A. S. and Lukin, M. D.},
  journal = {Phys. Rev. Lett.},
  volume = {92},
  issue = {23},
  pages = {230801},
  numpages = {4},
  year = {2004},
  month = {Jun},
  publisher = {American Physical Society},
  doi = {10.1103/PhysRevLett.92.230801}
}

@article{Giovannetti2006,
  title = {Quantum Metrology},
  author = {Giovannetti, Vittorio and Lloyd, Seth and Maccone, Lorenzo},
  journal = {Phys. Rev. Lett.},
  volume = {96},
  issue = {1},
  pages = {010401},
  numpages = {4},
  year = {2006},
  month = {Jan},
  publisher = {American Physical Society},
  doi = {10.1103/PhysRevLett.96.010401},
  url = {https://link.aps.org/doi/10.1103/PhysRevLett.96.010401}
}

@article{Shah2010,
  title = {High Bandwidth Atomic Magnetometery with Continuous Quantum Nondemolition Measurements},
  author = {Shah, V. and Vasilakis, G. and Romalis, M. V.},
  journal = {Phys. Rev. Lett.},
  volume = {104},
  issue = {1},
  pages = {013601},
  numpages = {4},
  year = {2010},
  month = {Jan},
  publisher = {American Physical Society},
  doi = {10.1103/PhysRevLett.104.013601},
  url = {https://link.aps.org/doi/10.1103/PhysRevLett.104.013601}
}

@article{Escher2011,
  title={General framework for estimating the ultimate precision limit in noisy quantum-enhanced metrology},
  author={Escher, BM and de Matos Filho, Ruynet Lima and Davidovich, Luiz},
  journal={Nature Physics},
  volume={7},
  number={5},
  pages={406--411},
  year={2011},
  publisher={Nature Publishing Group UK London},
  doi = {10.1038/nphys1958}
}

@article{Matsuzaki2011,
  title = {Magnetic field sensing beyond the standard quantum limit under the effect of decoherence},
  author = {Matsuzaki, Yuichiro and Benjamin, Simon C. and Fitzsimons, Joseph},
  journal = {Phys. Rev. A},
  volume = {84},
  issue = {1},
  pages = {012103},
  numpages = {5},
  year = {2011},
  month = {Jul},
  publisher = {American Physical Society},
  doi = {10.1103/PhysRevA.84.012103},
  url = {https://link.aps.org/doi/10.1103/PhysRevA.84.012103}
}

@article{Demkowicz2012,
  title={The elusive {H}eisenberg limit in quantum-enhanced metrology},
  author={Demkowicz-Dobrza{\'n}ski, Rafa{\l} and Ko{\l}ody{\'n}ski, Jan and Gu{\c{t}}{\u{a}}, M{\u{a}}d{\u{a}}lin},
  journal={Nature Communications},
  volume={3},
  number={1},
  pages={1063},
  year={2012},
  publisher={Nature Publishing Group UK London},
  doi = {10.1038/ncomms2067}
}

@article{Chin2012,
  title = {Quantum Metrology in Non-{M}arkovian Environments},
  author = {Chin, Alex W. and Huelga, Susana F. and Plenio, Martin B.},
  journal = {Phys. Rev. Lett.},
  volume = {109},
  issue = {23},
  pages = {233601},
  numpages = {5},
  year = {2012},
  month = {Dec},
  publisher = {American Physical Society},
  doi = {10.1103/PhysRevLett.109.233601},
  url = {https://link.aps.org/doi/10.1103/PhysRevLett.109.233601}
}

@article{Pospelov2013,
  title = {Detecting Domain Walls of Axionlike Models Using Terrestrial Experiments},
  author = {Pospelov, M. and Pustelny, S. and Ledbetter, M. P. and Kimball, D. F. Jackson and Gawlik, W. and Budker, D.},
  journal = {Phys. Rev. Lett.},
  volume = {110},
  issue = {2},
  pages = {021803},
  numpages = {5},
  year = {2013},
  month = {Jan},
  publisher = {American Physical Society},
  doi = {10.1103/PhysRevLett.110.021803},
  url = {https://link.aps.org/doi/10.1103/PhysRevLett.110.021803}
}

@article{Borregard2013,
  title = {Near-{H}eisenberg-Limited Atomic Clocks in the Presence of Decoherence},
  author = {Borregaard, J. and S\o{}rensen, A. S.},
  journal = {Phys. Rev. Lett.},
  volume = {111},
  issue = {9},
  pages = {090801},
  numpages = {5},
  year = {2013},
  month = {Aug},
  publisher = {American Physical Society},
  doi = {10.1103/PhysRevLett.111.090801},
}

@article{Kessler2014,
  title = {{H}eisenberg-Limited Atom Clocks Based on Entangled Qubits},
  author = {Kessler, E. M. and K\'om\'ar, P. and Bishof, M. and Jiang, L. and S\o{}rensen, A. S. and Ye, J. and Lukin, M. D.},
  journal = {Phys. Rev. Lett.},
  volume = {112},
  issue = {19},
  pages = {190403},
  numpages = {5},
  year = {2014},
  month = {May},
  publisher = {American Physical Society},
  doi = {10.1103/PhysRevLett.112.190403},
  url = {https://link.aps.org/doi/10.1103/PhysRevLett.112.190403}
}

@book{Ziemer2014,
  title={Principles of communications},
  author={Ziemer, Rodger E and Tranter, William H},
  year={2014},
  publisher={John Wiley \& Sons}
}

@article{Degen2017,
  title = {Quantum sensing},
  author = {Degen, C. L. and Reinhard, F. and Cappellaro, P.},
  journal = {Rev. Mod. Phys.},
  volume = {89},
  issue = {3},
  pages = {035002},
  numpages = {39},
  year = {2017},
  month = {Jul},
  publisher = {American Physical Society},
  doi = {10.1103/RevModPhys.89.035002},
  url = {https://link.aps.org/doi/10.1103/RevModPhys.89.035002}
}

@article{Pezze2018,
  title = {Quantum metrology with nonclassical states of atomic ensembles},
  author = {Pezz\`e, Luca and Smerzi, Augusto and Oberthaler, Markus K. and Schmied, Roman and Treutlein, Philipp},
  journal = {Rev. Mod. Phys.},
  volume = {90},
  issue = {3},
  pages = {035005},
  numpages = {70},
  year = {2018},
  month = {Sep},
  publisher = {American Physical Society},
  doi = {10.1103/RevModPhys.90.035005},
  url = {https://link.aps.org/doi/10.1103/RevModPhys.90.035005}
}

@article{Bing2019,
    author = {Gu, Bing and Franco, Ignacio},
    title = {When can quantum decoherence be mimicked by classical noise?},
    journal = {The Journal of Chemical Physics},
    volume = {151},
    number = {1},
    pages = {014109},
    year = {2019},
    month = {07},
    issn = {0021-9606},
    doi = {10.1063/1.5099499}
}

@article{Smorra2019,
  title={Direct limits on the interaction of antiprotons with axion-like dark matter},
  author={Smorra, C and Stadnik, YV and Blessing, PE and Bohman, M and Borchert, MJ and Devlin, JA and Erlewein, S and Harrington, JA and Higuchi, T and Mooser, A and others},
  journal={Nature},
  volume={575},
  number={7782},
  pages={310--314},
  year={2019},
  publisher={Nature Publishing Group UK London},
  doi={10.1038/s41586-019-1727-9}
}

@article{Stadnik2020,
  title = {Axion-induced effects in atoms, molecules, and nuclei: Parity nonconservation, anapole moments, electric dipole moments, and spin-gravity and spin-axion momentum couplings},
  author = {Stadnik, Y. V. and Flambaum, V. V.},
  journal = {Phys. Rev. D},
  volume = {89},
  issue = {4},
  pages = {043522},
  numpages = {14},
  year = {2014},
  month = {Feb},
  publisher = {American Physical Society},
  doi = {10.1103/PhysRevD.89.043522},
  url = {https://link.aps.org/doi/10.1103/PhysRevD.89.043522}
}

@article{Sedda2020,
  title={The missing link in gravitational-wave astronomy: discoveries waiting in the decihertz range},
  author={Sedda, Manuel Arca and Berry, Christopher PL and Jani, Karan and Amaro-Seoane, Pau and Auclair, Pierre and Baird, Jonathon and Baker, Tessa and Berti, Emanuele and Breivik, Katelyn and Burrows, Adam and others},
  journal={Classical and Quantum Gravity},
  volume={37},
  number={21},
  pages={215011},
  year={2020},
  publisher={IOP Publishing},
  doi={10.1088/1361-6382/abb5c1}
}

@article{Wan2022,
  title = {Bounds on adaptive quantum metrology under {M}arkovian noise},
  author = {Wan, Kianna and Lasenby, Robert},
  journal = {Phys. Rev. Res.},
  volume = {4},
  issue = {3},
  pages = {033092},
  numpages = {18},
  year = {2022},
  month = {Aug},
  publisher = {American Physical Society},
  doi = {10.1103/PhysRevResearch.4.033092}
}

@article{Colombo2022,
    author = {Colombo, Simone and Pedrozo-Peñafiel, Edwin and Vuletić, Vladan},
    title = {Entanglement-enhanced optical atomic clocks},
    journal = {Applied Physics Letters},
    volume = {121},
    number = {21},
    pages = {210502},
    year = {2022},
    month = {11},
    issn = {0003-6951},
    doi = {10.1063/5.0121372},
}

@article{Polloreno2023,
  title = {Opportunities and Limitations in Broadband Sensing},
  author = {Polloreno, Anthony M. and Beckey, Jacob L. and Levin, Joshua and Shlosberg, Ariel and Thompson, James K. and Foss-Feig, Michael and Hayes, David and Smith, Graeme},
  journal = {Phys. Rev. Appl.},
  volume = {19},
  issue = {1},
  pages = {014029},
  numpages = {14},
  year = {2023},
  month = {Jan},
  publisher = {American Physical Society},
  doi = {10.1103/PhysRevApplied.19.014029}
}

@article{Xia2023,
  title={Entanglement-enhanced optomechanical sensing},
  author={Xia, Yi and Agrawal, Aman R and Pluchar, Christian M and Brady, Anthony J and Liu, Zhen and Zhuang, Quntao and Wilson, Dalziel J and Zhang, Zheshen},
  journal={Nature Photonics},
  volume={17},
  number={6},
  pages={470--477},
  year={2023},
  publisher={Nature Publishing Group UK London},
  doi={10.1038/s41566-023-01178-0}
}

@article{Ye2024,
  title = {Essay: Quantum Sensing with Atomic, Molecular, and Optical Platforms for Fundamental Physics},
  author = {Ye, Jun and Zoller, Peter},
  journal = {Phys. Rev. Lett.},
  volume = {132},
  issue = {19},
  pages = {190001},
  numpages = {11},
  year = {2024},
  month = {May},
  publisher = {American Physical Society},
  doi = {10.1103/PhysRevLett.132.190001},
  url = {https://link.aps.org/doi/10.1103/PhysRevLett.132.190001}
}

@article{Boyers2025,
  title = {Spin squeezing of macroscopic nuclear spin ensembles},
  author = {Boyers, Eric and Goldstein, Garry and Sushkov, Alexander O.},
  journal = {Phys. Rev. D},
  volume = {111},
  issue = {5},
  pages = {052004},
  numpages = {16},
  year = {2025},
  month = {Mar},
  publisher = {American Physical Society},
  doi = {10.1103/PhysRevD.111.052004},
  url = {https://link.aps.org/doi/10.1103/PhysRevD.111.052004}
}

@article{Gosh2025,
  title={Optimal phase estimation in the presence of correlated dephasing},
  author={Ghosh, Srijon and Kobus, Arkadiusz and Kurdzia{\l}ek, Stanis{\l}aw and Demkowicz-Dobrza{\'n}ski, Rafa{\l}},
  journal={arXiv preprint arXiv:2511.07211},
  year={2025}
}

@article{Sichanugrist2025,
  title = {Entanglement-enhanced ac magnetometry in the presence of {M}arkovian noise},
  author = {Sichanugrist, Thanaporn and Fukuda, Hajime and Moroi, Takeo and Nakayama, Kazunori and Chigusa, So and Mizuochi, Norikazu and Hazumi, Masashi and Matsuzaki, Yuichiro},
  journal = {Phys. Rev. A},
  volume = {111},
  issue = {4},
  pages = {042605},
  numpages = {11},
  year = {2025},
  month = {Apr},
  publisher = {American Physical Society},
  doi = {10.1103/PhysRevA.111.042605}
}

@article{kaufmann2026,
  title={Entanglement Enhanced Sensing with Qubits affected by non-{M}arkovian Dephasing},
  author={Kaufmann, Noah and Nielsen, Kasper H and Gonz{\'a}lez-Ruiz, Eva M and S{\o}rensen, Anders S},
  journal={arXiv preprint arXiv:2604.22368},
  year={2026}
}

@article{polloreno2026,
  title={Simple broadband signal detection at the fundamental limit},
  author={Polloreno, Anthony M and Smith, Graeme},
  journal={arXiv preprint arXiv:2601.19816},
  year={2026}
}

\end{document}